\documentclass [aps,prx,twocolumn,groupeauthor,showpacs,superscriptaddress]{revtex4}
\usepackage{amsmath,amssymb,latexsym}

\usepackage{graphicx}
\usepackage{bm}

\newcommand{\JJ}{J_\text{ex}}

\begin{document}

\title{Non-thermal melting of N\'eel order in the Hubbard model}
\author{Karsten~Balzer}
\affiliation{Max Planck Research Department for Structural Dynamics, University of Hamburg-CFEL, 22761 Hamburg, Germany}
\author{F. Alexander Wolf}
\affiliation{
Department of Physics,
Arnold Sommerfeld Center for Theoretical Physics,
LMU Munich,
Theresienstrasse 37,
80333 M\"unchen, Germany}	
\author{Ian P. McCulloch}
\affiliation{
Centre for Engineered Quantum Systems,
School of Physical Sciences,
The University of Queensland,
Brisbane, Queensland 4072, Australia}
\author{Philipp~Werner}
\affiliation{Department of Physics, University of Fribourg, 1700 Fribourg, Switzerland}
\author{Martin~Eckstein}
\email{martin.eckstein@mpsd.cfel.de}
\affiliation{Max Planck Research Department for Structural Dynamics, University of Hamburg-CFEL, 22761 Hamburg, Germany}

\begin{abstract}
We study the unitary time evolution of antiferromagnetic order in the Hubbard model after a quench starting from the perfect N\'eel state. 
In this setup, which is well suited for experiments with cold atoms, one can distinguish fundamentally different pathways for melting of long-range order at weak and strong interaction. In the Mott insulating regime, melting of long-range order occurs due to the ultra-fast transfer of energy from charge excitations to the spin background, while local magnetic moments and their exchange coupling persist during the process. The latter can be demonstrated by a local spin-precession experiment. At weak interaction, local moments decay along with the long-range order. The dynamics is governed by residual quasiparticles, which are reflected in oscillations of the off-diagonal components of the momentum distribution. Such oscillations provide an alternative route to study the prethermalization phenomenon and its influence on the dynamics away from the integrable (noninteracting) limit. The Hubbard model is solved within nonequilibrium dynamical mean-field theory, using the density matrix-renormalization group as an impurity solver.
\end{abstract}
\date{\today}
\maketitle

\section{Introduction}

Ultra-fast pump-probe experiments on condensed matter systems and experiments with cold gases in optical lattices have opened the intriguing possibility of controlling transitions between complex phases on microscopic timescales. This has motivated intensive theoretical efforts to understand fundamental aspects of the dynamics in interacting many-body systems, and leads to predictions in marked contrast to the naive expectation that interactions imply rapid  thermalization  \cite{RevModPhys.83.863}: Integrable systems can keep memory of the initial state for all times and relax to a generalized Gibbs ensemble \cite{PhysRevLett.98.050405, Langen2014}, but also away from integrability thermalization can be delayed by prethermalizaton \cite{Berges2004, Moeckel2008, Kollar2011, Gring2012}, and one can identify regimes of different dynamical behavior which are clearly separated by non-thermal critical points \cite{Kollath2007, Eckstein2009, Barmettler2009, Schiro2010, PhysRevLett.105.220401, Tsuji2013}. 

Of particular interest with respect to complex phases in condensed matter is the dynamics of symmetry broken-states \cite{Beaud2014, Ehrke2011, Ichikawa2011}. While the relevant relaxation mechanisms after a perturbation 
are hard to disentangle in a solid, cold atoms in optical lattices provide a versatile platform to investigate isolated quantum systems in ideal situations. The preparation of thermodynamic long-range ordered phases in cold atoms is still a challenge \cite{Greif14062013}, but advanced techniques for lattice design have made it possible to prepare an ordered state on a lattice of isolated sites, and to probe its dynamics after 
tunneling between the sites is switched on \cite{Trotzky2012, Schreiber2015, PhysRevLett.113.170403, Brown2014}. In the following we consider such a setup for the Fermi-Hubbard model, a paradigm model for emergent long-range order in condensed matter systems. 
We will simulate the time evolution starting from a classical N\'eel state in which neighboring lattice sites of a bipartite lattice are occupied with particles of opposite spin. 

In general, one can anticipate fundamentally different pathways for melting of long-range antiferromagnetic order in the weakly and strongly interacting Hubbard model: For strong interaction, long-range order arises from antiferromagnetically coupled local moments, which emerge when charge fluctuations are frozen. Magnetic order could thus possibly melt via the destruction of the local moments themselves, through a reduction of the effective exchange interaction \cite{Mentink2014a} (while moments persist), or, along a quasi-thermal pathway, by the transfer of energy from excited quasiparticles (hot electrons) to spins. The latter mechanism is intensively studied in the context of photo-carrier relaxation in high-T$_c$ cuprates \cite{Mierzejewski2011, Werner2012, Kogoj2014, Eckstein2014, Eckstein2014a, Golez2014}, where the investigation of the spin-charge interaction challenges  the limits for the time-resolution in state of the art pump-probe experiments \cite{Okamoto2010, Okamoto2011, DalConte2015}. For weak interaction, on the other hand, quasiparticle states may be important to understand relaxation processes. In the paramagnetic phase the conservation of the quasiparticle momentum occupations imposes constraints on the dynamics which can lead to prethermalization \cite{Berges2004, Moeckel2008, Eckstein2009, Hamerla2013, Hamerla2014, Tsuji2014}. Prethermalizaton, which was recently observed in a one-dimensional Bose gas \cite{Gring2012}, has been suggested to be a universal feature of near-integrable systems \cite{Kollar2011}, but previous predictions for the Hubbard model rely on a discontinuity of the momentum distribution which is absent at nonzero temperature and thus experimentally hard to observe. Here we show that the symmetry-broken initial state provides an alternative perspective to investigate this physics and its breakdown far from integrability.

Quenches from a N\'eel state have been explored in quantum spin models \cite{Barmettler2009, Barmettler2010, Liu2014, Heyl2014}, also as a way to prepare ordered states in the Hubbard model \cite{Ojekhile2013}, but a pure spin model cannot describe the relevant dynamics of charge excitations and local moments. The Hubbard model has been studied in one dimension using 
the density-matrix renormalization group (DMRG) 
\cite{A.BauerF.Dorfner2015}. For the dynamics of lattice fermion models in more than one dimension, nonequilibrium dynamical mean-field theory (DMFT)  \cite{Aoki2014} is the most promising approach. Quenches within the antiferromagnetic phase of the Hubbard model at strong-coupling \cite{Werner2012} are in line with the ``quasi-thermal'' pathway discussed above. The regime of intermediate interactions, where the notion of local moments becomes ambiguous, or to weak coupling, where prethermalization may be expected, has been elusive so far. Previous numerical solutions of the DMFT equations were based on the self-consistent strong-coupling expansion \cite{Eckstein2010} or weak-coupling impurity solvers~\cite{Eckstein2011a, Tsuji2013, Tsuji2013weakcouplingprb}, which both fail at intermediate coupling,  while weak-coupling quantum Monte Carlo studies \cite{Eckstein2009,Eckstein2010} are most efficient for noninteracting initial states and restricted to short times. In this work we overcome these limitations using a recently developed Hamiltonian based formulation for the impurity model of nonequilibrium DMFT \cite{Gramsch2013}, which has opened the possibility to use wave-function based techniques to solve the DMFT equations \cite{Balzer2015,Wolf2014}. Here we use
DMRG as an impurity solver 
\cite{Wolf2014}, which allows us to reach sufficiently long times in the evolution to address the above issues. 

\section{Model and methods} 

Throughout  this work  we consider the single-band Hubbard model at half-filling, with nearest-neighbor hopping 
$J$ and on-site Coulomb repulsion $U$. The Hamiltonian is given by
\begin{equation}
\label{hubbard}
H = 
- 
J(t)
\sum_{\langle ij \rangle \sigma=\uparrow,\downarrow}
c_{i\sigma}^\dagger c_{j\sigma}
+
U \sum_{i} 
(n_{i\uparrow}-\tfrac12)
(n_{i\downarrow}-\tfrac12),
\end{equation}
where $c_{i\sigma}^\dagger$ ($c_{i\sigma}$) are electron creation (annihilation) operators for lattice site $i$ and spin $\sigma$, and $n_{i\sigma} = c_{i\sigma}^\dagger c_{i\sigma}$. The model is solved using nonequilibrium DMFT \cite{Aoki2014}, for a Bethe lattice in the limit of infinite coordination number $Z$ and hopping $J=J_*/\sqrt{Z}$, where the approach becomes exact \cite{PhysRevLett.62.324}. The energy unit is set by $J_*=1$, and time is measured in inverse energy, i.e, the free density of states is given by $D(\epsilon) = \sqrt{4-\epsilon^2}/(2\pi)$.  To simulate the quench, we choose a time-dependent hopping $J_*(t)=0$ for $t\le 0$ and $J_*(t)=1$ for $t>0$. For $t\le 0$, the system therefore consists of a set of isolated lattice sites, which are prepared in a classical  N\'eel state,  
\begin{align}
\label{neel}
|\Psi_\text{N\'eel}\rangle = \prod_{i\in A} c_{i\uparrow}^\dagger \prod_{j\in B} c_{j\downarrow}^\dagger |0\rangle,
\end{align}
where $A$ and $B$ are sub-lattices of the bipartite Bethe lattice. 

In DMFT, the lattice model is mapped to a set of impurity problems, one for each inequivalent lattice site $j=A,B$, with a time-dependent hybridization function $\Delta_{j\sigma}(t,t')$. (In this expression, time arguments lie on the Keldysh contour; see Ref.~\cite{Aoki2014} for a detailed description of nonequilibrium DMFT and the Keldysh formalism.) For the Bethe lattice, the latter is determined self-consistently by $\Delta_{A(B),\sigma}(t,t') = J_*(t)G_{B(A),\sigma}(t,t')J_*(t')$, where $G_{j\sigma}(t,t') = -i \langle \text{T}_\mathcal{C} c_{j\sigma}(t)c_{j\sigma}^\dagger(t')\rangle$ is the local Green's function. To solve the impurity model with a non time-translationally invariant hybridization function, we derive an equivalent representation in terms of a time-dependent Anderson impurity Hamiltonian \cite{Gramsch2013} with up to $L=24$ bath orbitals, from which the time-dependent Green's functions are computed using a Krylov time-propagation for matrix product states \cite{Wolf2014}. The Hamiltonian representation of the DMFT impurity model is exact for small times, but an increasing number of bath sites is needed to reach longer times \cite{Balzer2014}. We verify the convergence of the solution with the bath size $L$. Up to $L=12$, the results have also been cross checked with a Krylov time-propagation in the full Hilbert space. For further details of the numerical solution, see Appendix~\ref{appDMFT}.

\section{Results} 

Figure \ref{fig1} shows the time evolution of the antiferromagnetic order parameter $M(t)$
and the double occupation $d(t)=\langle n_{\uparrow}(t)n_{\downarrow}(t)\rangle$ after the quench, 
for various values of the Coulomb interaction. In order to account for the trivial reduction of 
the local spin expectation value by virtual charge fluctuations, we define $M(t)$ as the staggered order 
$M_\text{stagg}
\equiv
\langle n_{A\uparrow}(t)-n_{A\downarrow}(t)\rangle$
=
$\langle n_{B\downarrow}(t)-n_{B\uparrow}(t)\rangle$,  normalized by the probability $P_1$ for a site to be singly occupied, $P_1(t)=1-2d(t)$.  To test whether the system thermalizes after the quench, we compare to an equilibrium state at the same internal energy (which is zero for the N\'eel state). The corresponding effective temperature $T_\text{eff}$ (Fig.~\ref{fig1}c) lies above the N\'eel temperature $T_\text{N\'eel}$ for all values of $U$  \cite{footnote2}.
This implies a paramagnetic state after thermalization. While $M(t)$ indeed continues to decay throughout the simulated time interval, the double occupancy saturates to a non-thermal value for $U\gtrsim4$ (arrows in Fig.~\ref{fig1}b point at the thermalized value  $d(T_\text{eff})$), in agreement with earlier studies on the lifetime of doublons in the paramagnetic Mott regime \cite{Winkler2006, Sensarma2010, Eckstein2011}. Already at a first glance, the relaxation of $M(t)$ and $d(t)$ therefore suggests different mechanisms for small and large values of $U$, with a rapid and oscillatory decay of $M(t)$, and a long-lived non-thermal state, respectively. In the following we will analyze the two regimes in more detail.

\begin{figure}[tbp]
\begin{center}
\includegraphics[angle=0, width=\columnwidth]{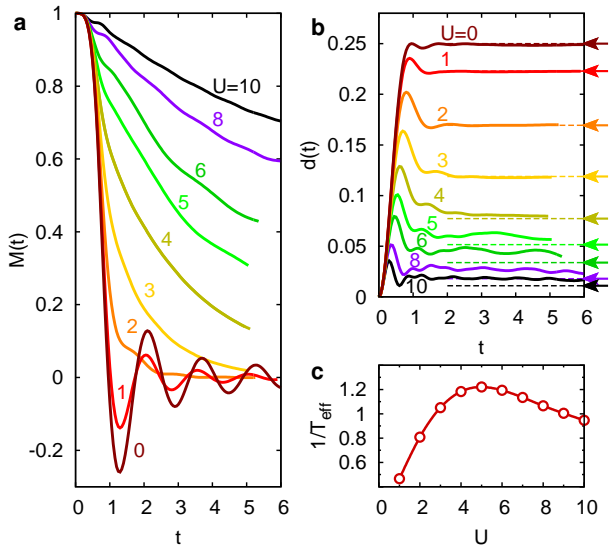}%
\caption{
{\bf a)} Time evolution of the order parameter $M(t) = M_\text{stagg}/[1-2d(t)]$ for 
different values of the Coulomb repulsion in the range $U=0$ to $U = 10$.
{\bf b)} The double occupation $d(t)$ at the same values of $U$. Arrows indicate 
the double occupation in a thermalized state at the same total energy as the quenched 
state (as obtained from equilibrium DMFT, using Continuous-time Quantum Monte 
Carlo \cite{Werner2006} as an impurity solver). All thermalized states are in the 
paramagnetic phase, the corresponding inverse temperature $1/T_\text{eff}$ 
is plotted in {\bf c)}.}
\label{fig1}
\end{center}
\end{figure}

\subsection{Weak-coupling: Residual quasiparticles} 

\begin{figure*}[tbp]
\begin{center}
\includegraphics[angle=0, width=0.8\textwidth]{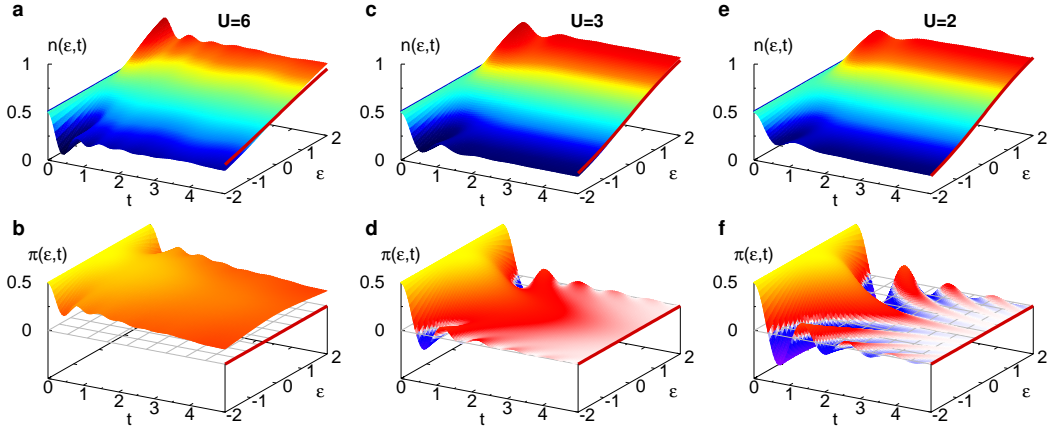}%
\caption{ Diagonal component $n(\epsilon_k, t)=\langle c_k^\dagger c_{k}\rangle$ of the momentum occupation (panels {\bf a, c}, and {\bf e}) and off-diagonal component  $\text{Re}\, \pi(\epsilon_k) = \langle c_k^\dagger c_{\bar k}\rangle$ (panels {\bf b, d}, and {\bf f}), where $k$ and $\bar k$ are pairs of single particle states coupled by a staggered potential, plotted for quenches to three different values of $U$ as a function of the energy $\epsilon_k$ ranging from $-2$ to $2$ in the band of the Bethe lattice. The bold lines indicate momentum distributions obtained in the (paramagnetic) equilibrium state at the same energy.}
\label{fig2}
\end{center}
\end{figure*} 

For quenches to small $U$ the Hamiltonian is close to the integrable point $U=0$. This suggests to study relaxation in terms of the momentum occupation $n_k(t)=\langle c_{k}^\dagger c_k\rangle$, which is conserved at $U=0$. For a state with translational symmetry breaking, the single-particle density matrix $\rho_{kk'}(t) = \langle c_{k'}^\dagger(t) c_{k}(t) \rangle$ is no longer diagonal in momentum $k$. (The discussion holds for a general lattice like the Bethe lattice when $k$ denotes the eigenstates of the translationally invariant hopping matrix.) For nearest neighbor hopping on a bipartite lattice, eigenstates come in pairs $k,\bar k$ with single particle energy $\epsilon_{k}=-\epsilon_{\bar k}$, where the wave functions for $\bar k$ and $k$ differ by a staggered phase $\xi_i = \pm$ for $i\in A(B)$, and $\rho_{k\bar k} \neq 0$ if the symmetry between sub-lattices is broken. (On the cubic lattice, $k$ and $\bar k = k+(\pi,\pi,\ldots)$ are  momenta related by the antiferromagnetic nesting vector.) In Fig.~\ref{fig2} we plot the diagonal and off-diagonal components of the single-particle density matrix in terms of the two functions $n(\epsilon_k, t) = \langle c_{k}^\dagger(t) c_{k}(t) \rangle$ and $\pi(\epsilon_k, t) = \langle c_{k}^\dagger(t) c_{\bar k}(t) \rangle$, which depend on $k$ only via $\epsilon_k$ due to the locality of the self-energy within DMFT. In the thermalized state, $\pi(\epsilon)=0$ because the state does not break the sub-lattice symmetry, while in the localized initial state, $n(\epsilon, t=0)=\pi (\epsilon, t=0)=1/2$. In agreement with the behavior of the double occupancy,  $n(\epsilon, t)$ does not thermalize at large $U$ (thermalized values $\pi_{T_\text{eff}}(\epsilon)$ and $n_{T_\text{eff}}(\epsilon)$ are shown by solid lines). For $U\le 2$, however, differences between $n(\epsilon,t)$ and $n_{T_\text{eff}}(\epsilon)$ become tiny. This is in stark contrast to the behavior of the paramagnetic system after a quench from $U=0$, where prethermalization manifests itself precisely in the difference between $n(\epsilon,t)$ and $n_{T_\text{eff}}(\epsilon)$ \cite{Moeckel2008}. Moreover, around $U=3$ the relaxation of $\pi(\epsilon,t)$ changes from an oscillatory to a  monotonous decay. (In Fig.~\ref{fig2}, we plot the real part of $\pi(\epsilon,t)$, the imaginary part shows a similar crossover from oscillatory to non-oscillatory behavior.)

\begin{figure}[tbp]
\begin{center}
\includegraphics[angle=0, width=0.9\columnwidth]{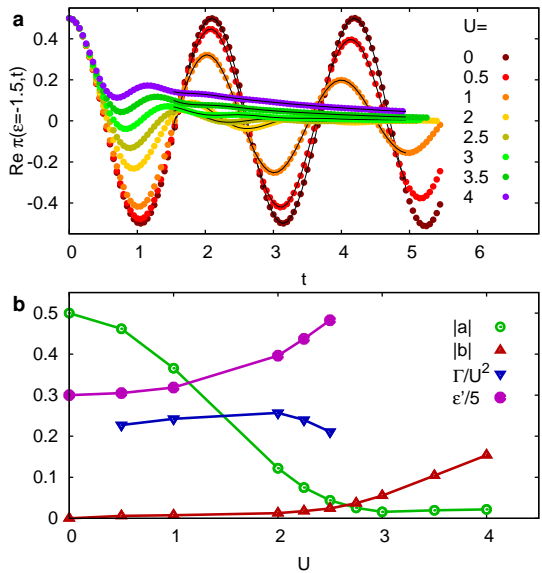}
\caption{ {\bf a)} Line-out of $\pi(\epsilon,t)$ for $\epsilon=-1.5$ and various values of $U$. Solid lines are fits with the sum of a decaying exponential background and decaying  oscillations, $f(t)=b\exp(-c(t-t_0)) + a \exp(-\Gamma (t-t_0))\cos(2\epsilon' t+\phi)$ for $t>t_0=1.5$. {\bf b)} Amplitudes of the background, $b$, the oscillations, $a$, the quasi-particle energy $\epsilon'$ and the quasi-particle decay rate $\Gamma$ as a function of $U$.}
\label{fig3}
\end{center}
\end{figure}

This observation may be explained following the perturbative arguments of Refs.~\cite{Moeckel2008,Kollar2011}. To second order in $U$, the Hamiltonian \eqref{hubbard} is unitarily equivalent to a model $H = \sum_{k} \tilde \epsilon_k  \tilde c_{k}^\dagger \tilde c_{k} + \mathcal{O}(U^2)$ which is quadratic in terms of quasiparticle operators $\tilde c_k$; we have  $c_k = R_k \tilde c_k + incoh.$ with a finite residue $R_k$, where $incoh.$  denotes incoherent contributions, i.e., an admixture of particle-hole excitations to higher order in $U$. Hence the momentum occupation is given by $n_{k}(t) = R_k^2 \langle \tilde c_{k}^\dagger(t) \tilde c_{k}(t)  \rangle + incoh. $. The term proportional to $R_k^2$ (the coherent part) is unchanged by the time evolution to second order in $U$. Back-transforming gives $n_k(t) = R_k^4 n_k(0) + incoh.$, where the incoherent contribution is a smooth function of $k$. For quenches in the paramagnetic phase, $n_k(t)$ thus preserves the initial discontinuity at the Fermi surface, which can be taken as a measure of prethermalization \cite{Moeckel2008}. In the symmetry broken state, however, $n_k(0)$ is independent of $k$, and thus $n_k(t)$ does not clearly exhibit the existence of residual quasiparticles. In fact, the numerical results suggest that the incoherent part can accurately be described by a thermal distribution. In contrast, a similar argument for the off-diagonal component shows that  $\pi_{k}(t) = R_k^2   \langle \tilde c_{k}^\dagger(t) \tilde c_{\bar k}(t)  \rangle + incoh. = R_k^4 e^{i 2 \tilde \epsilon_k t}  \pi(\epsilon,t=0)  + incoh. $, where we have used the time evolution of the quasiparticle, $\tilde c_{k(\bar k)} (t) = e^{\mp i\tilde \epsilon_k t} \tilde c_{k(\bar k)} (0)$ and $R_k=R_{\bar k}$. Hence we find that the residual quasiparticle dynamics leading to prethermalization close to the integrable point $U=0$ can be studied very conveniently with the symmetry broken initial state in terms of oscillations in the off-diagonal components of the momentum occupation.

Similar to the interaction quench in the paramagnetic phase \cite{Eckstein2009,Schiro2010}, we find that the ``prethermalization'' regime in which residual quasiparticles dominate the dynamics is limited to small interactions; at large interactions, $\pi(\epsilon,t)$ relaxes to zero monotonously (see the $U=6$ data in Fig.~\ref{fig2}b) and the distribution becomes flat over the Brillouin zone. Below we will see that the dynamics at large $U$ can be analyzed in terms of well defined localized moments. In contrast to the quench in the paramagnetic phase, the crossover between the weak and strong coupling regimes is relatively smooth and occurs between $U=2$ and $U=3$: In Fig.~\ref{fig3}a, we exemplarily plot $\text{Re}\,\pi(\epsilon,t)$ for fixed $\epsilon=-1.5$ and various $U$. For $U\lesssim 2$, the curves can be accurately fit with decaying oscillations $f_1(t)=a\exp(-\Gamma t)\cos(-2\epsilon' t+\phi)$, where in agreement with the discussion above the quasi-particle energy  $\epsilon'\to \epsilon$, and $\Gamma \sim U^2$ for $U\to 0$ (solid lines in Fig.~\ref{fig3}a, fit parameters in Fig.~\ref{fig3}b). For $U\gtrsim 3$, on the other hand, a good fit is a monotonously decaying curve $f_2(t)=b\exp(-c t)$. For $2\lesssim U \lesssim 3$, there is a crossover between the two behaviors, as evidenced by the dependence of the amplitudes $a$ and $b$ of the monotonous and the oscillating component on $U$ (Fig.~\ref{fig3}b). 

Before discussing the strong-coupling regime, we note that off-diagonal momentum distributions can in principle be measured by a modified time-of-flight measurement, if before releasing the cloud, one would switch off the tunneling and the interaction, switch on a staggered potential which is $\pm \Delta$ on the A and B sub-lattice, respectively, and evolve for a given time $t_m$. Time of flight measures the regular momentum occupation $n_{k}=\langle c_k^\dagger  c_k \rangle$ after that procedure. In the $k,\bar k$ basis, the staggered potential is given by $H= \Delta \sum_{k} (c_{k}^\dagger c_{\bar k} + H.c.)$, so that $n_{k}(t+t_m) = n_{k}(t) \cos^2(t_m\Delta ) n_{\bar k}(t)+ \sin^2(t_m\Delta )   + \sin(2t_m \Delta  ) \text{Im}\, \pi_{k}(t)$ after propagation in the pure staggered potential from time $t$ to $t+t_m$, and $\text{Im}\,\pi_{k}(t)$ can be extracted.

\subsection{Dynamics of local moments} 

\begin{figure*}[tbp]
\begin{center}
\includegraphics[angle=0, width=\textwidth]{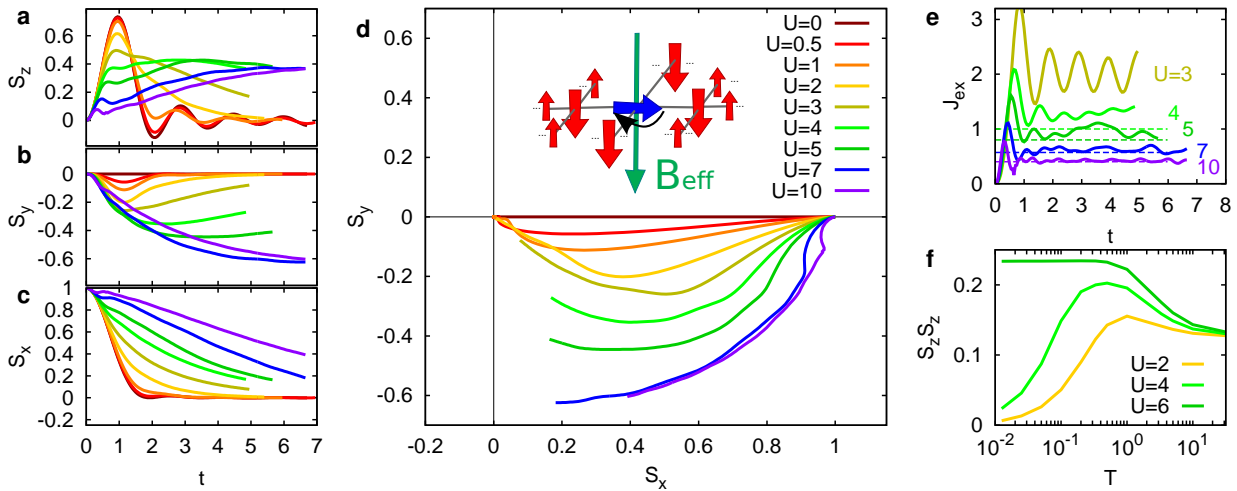}%
\caption{
Dynamics at a probe site $o$, where the spin is initially flipped to the $x$-direction. {\bf a)}-{\bf c)} Expectation values $\langle S_z(t)\rangle$,  $\langle S_y(t)\rangle$, and $\langle S_x(t)\rangle$ for various values of $U$ (see legend in {\bf d)}). {\bf d)} Trajectory of the spin $\langle \bm S \rangle$ in the $S_x$-$S_y$ plane. The inset illustrates the initial state, with one spin flipped to the $x$ direction, and the effective exchange field. {\bf d)} Effective exchange interaction, $d\phi(t)/dt / |M_\text{stagg}|$, where $\phi(t)=\text{atan}(S_y/S_x)$ is the angle in the  $S_x$-$S_y$ plane, for $U=3,4,5,7,10$. The dotted lines correspond to the perturbative value of the exchange interaction $J_\text{ex}=4J_*^2/U.$ {\bf f)} Temperature-dependent local moment in equilibrium, defined by $(1/\beta) \int_0^\beta d\tau \langle S_z(\tau) S_z(0)\rangle$, obtained using CTQMC as an impurity solver.
}
\label{fig4}
\end{center}
\end{figure*} 

In a Mott insulator at large $U$ one can expect the existence of well-defined local moments. It is an intriguing question whether these moments persist in the quenched state while the long-range order disappears, and to what extent the crossover in relaxation behavior from weak to strong coupling can be characterized in terms of these local moments. In the following, we propose a simple experiment to distinguish the existence and strength of moments in the quenched state: one spin in the initial N\'eel state on a given site (the probe site ``$o$'') is flipped to the $x$-direction (see Fig.~\ref{fig4}d, inset). Choosing $o$ on the $A$-sublattice of the N\'eel state, the initial state \eqref{neel} of the dynamics is changed to $ (|\Psi_\text{N\'eel},\uparrow\rangle +|\Psi_\text{N\'eel},\downarrow\rangle )/\sqrt{2}$, where $|\Psi_\text{N\'eel},\sigma\rangle = c_{o,\sigma}^\dagger c_{o\uparrow} |\Psi_\text{N\'eel}\rangle  $. In a perfect local moment picture, the spin should then precess in the exchange field of it's neighbors. 

The inhomogeneous setup with one probe spin can be solved within DMFT, where it corresponds to a modified impurity problem at site $o$, while the rest of the lattice is unchanged (see App.~\ref{appDMFT}).  Figure~\ref{fig4}a-c shows the local spin expectation values $\langle S_{z}\rangle$, $\langle S_{y}\rangle$, and  $\langle S_{x}\rangle$ at site $o$ for various values of the interaction. In Fig.~\ref{fig4}d we show the trajectory of the spin in the $S_x$-$S_y$ plane, starting from $S_x=1$, $S_y=0$ at time $t=0$. For large $U$ one can indeed observe a precessional motion in the $S_x$-$S_y$ plane, as expected for a local moment subject to an exchange field in the $z$-direction. For $U=0$, on the other hand, the spin-dynamics is entirely longitudinal, showing no sign of well-defined local moments. (For $U=0$, the dynamics can be solved analytically, yielding $S_{x,o}=J_1(t)^2/t^2$, while $S_{o,y}=0$ for the Bethe lattice at $Z=\infty$, where $J_1(x)$ is the first Bessel function (Appendix~\ref{appU0})). There is a crossover between the two relaxation regimes. 

Although the exchange interaction is in principle not an instantaneous interaction on the timescale of the electronic hopping \cite{Mentink2014a}, it is illustrative to quantify the precession dynamics in terms of an effective exchange field. For this purpose we follow Refs.~\cite{Mentink2014a,Mentink2014} and define $\bm B_\text{eff}$ such that  $\langle \bm S (t)\rangle$ satisfies the equation of motion $d/dt \langle \bm S (t)\rangle = \bm B_\text{eff} \times \langle \bm S (t)\rangle$. We can assume that $\bm B_\text{eff}=B_\text{eff}\hat{\bm z}$ acts only in the $z$ direction (parallel to the order parameter $ M_\text{stagg}$ on the neighboring sites), and use the parametrization $ B_\text{eff} = J_\text{ex} M_\text{stagg}$ to define an effective exchange interaction  $J_\text{ex}$; the latter is then given by $J_\text{ex} = \dot\phi(t) / |M_\text{stagg}|$ where $\phi(t)=\text{atan}(S_y(t)/S_x(t))$ is the angle of the spin in the $x$-$y$ plane. The resulting value $J_\text{ex}$ is plotted in Fig.~\ref{fig4}e. For large $U$, $J_\text{ex}$ shows very good agreement with the perturbative value of the exchange in the Hubbard model, $4J_*^2/U$, and is not substantially decreasing with time even for quenches at intermediate interaction ($U\approx4$) where the order parameter quickly decays to zero (see Fig.~\ref{fig1}). Equilibrium estimates of the local moment in the intermediate coupling regime (Fig.~\ref{fig4}f) furthermore show tendencies of moment formation at elevated temperatures, which may explain why some spin precession occurs even for $U=2$. The combination of these results shows that the melting of long-range order proceeds by the ``quasi-thermal'' pathway discussed in the introduction, i.e., a disordering of exchange-coupled moments, rather than by 
a change of the exchange interaction or a destruction of the moments.

\begin{figure}[tbp]
\begin{center}
\includegraphics[angle=0, width=\columnwidth]{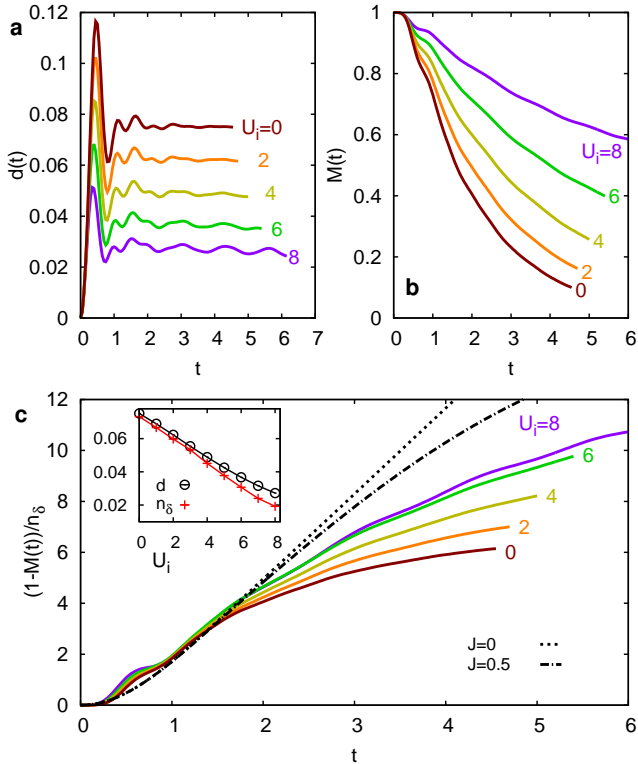}%
\caption{
(Color online) Data for the following quench protocol:
$t<0:$ $J_*=0$ (N\'eel state);
$0\le t <0.5$: $J_*=1$, $U=U_i$;
$t\ge 0.5$: $J_*=1$, $U=8$. The intermediate step controls the 
excitation density in the final state. 
{\bf a)} Time-evolution of the double occupancy or various values
of $U_i$. {\bf b)} Time-evolution of the order parameter $M(t)$.
{\bf c)} Number of spin-flips per charge carrier density $n_\delta$, $[1-M(t)]/n_\delta$, 
compared to the mean displacement $R(t)$ of the initially localized particle in the $t$-$J_z$ 
model  for $\JJ=0,0.5$  and $\JJ=0$ (dotted and dashed black lines, see text). The inset shows the the average of $d(t)$ 
for $2\le t\le 5$ (open circles), and the density of mobile carriers $n_\delta$, obtained from 
the integrated weight in the upper Hubbard band (red crosses).
}
\label{fig5}
\end{center}
\end{figure}

\subsection{Strong coupling: spin-charge interaction} 

At large $U$, a quench within a Mott insulator freezes virtual charge fluctuations, leaving behind a certain density 
$n_\delta$
of long-lived mobile carriers \cite{Werner2012}. The mechanism for the decay of the antiferromagnetic order is thus expected to be the transfer of energy from excited quasiparticles to the spins, which is currently intensively investigated in condensed matter pump-probe experiments. Although this mechanism is rather well understood in contrast to the dynamics at intermediate coupling, it is worthwhile to see how it can be investigated in the cold atom setup, because experiments in solids are very challenging. 

To investigate the decay of long-range order systematically, one has to vary the excitation density. Here we use a quench protocol where in addition to switching on the hopping at time $t=0$, the interaction is changed to an intermediate interaction value $U_i$ for a short time $0\le t\le 0.5$, before it is set to the final value $U$ for $t>0.5$.  (Note that various other protocols, such as an intermediate time-dependent modulation of the hopping, would have the same effect.)  Small values $U_i$ lead to a larger double occupancy (Fig.~\ref{fig5}a), and indeed also a more rapid decay of $M(t)$ (Fig.~\ref{fig5}b). We also note that an exponential fit $M(t) \sim ae^{-t/\tau}+b$ would be consistent with a threshold behavior in which  $M(t)$ extrapolates to a finite value $b$ for small excitation density ($U_i$ close to $U_i=8$) and to $b=0$ for large excitation density, consistent with earlier quench studies 
based on the non-crossing approximation impurity solver
\cite{Werner2012}, but the times are not sufficient to analyze this long-time behavior in detail.  

For a quantitative analysis of the short time behavior,  we determine the number $n_\delta$ of doublons and hole carriers in the quenched state (Fig.~\ref{fig5}c inset). Due to virtual charge fluctuations, $n_\delta$ is not exactly given by an instantaneous expectation value $d(t)$ in the Hubbard model, and we compute $n_{\delta}$ from the total weight in the upper Hubbard band (App.~\ref{appndoublon}) \cite{footnote1}. For small times, the curves $M(t)$ for various values $U_i$ can then be scaled on top of each other by plotting $[1-M(t)]/n_\delta$ (Fig.~\ref{fig5}c). Such a scaling implies that the number of flipped spins, $1-M(t)$, is proportional to the number of carriers. This is consistent with the picture that spin-flips are inserted by mobile carriers, which are initially localized and thus act independently up to times depending on $n_\delta$. For large times there is a deviation from the scaling due to the gradual melting of the order parameter. 

To further corroborate this picture, we analytically  compute the spin-flip rate per carrier in the low density limit from the behavior of a single carrier which is initially localized at a given site $0$ in a Ising spin background. Following Ref.~\cite{Golez2014}, we omit the transverse dynamics of the spins, which is on the timescale of $\JJ$ and much slower than the hopping, and keep only the $z$-component of the exchange coupling $\JJ$ ($t$-$J_z$ model). The model can then be reduced to a tight-binding model for a single particle on the lattice, with effective Hamiltonian $H = -J_*/\sqrt{Z}\sum_{\langle ij \rangle } c_{i}^\dagger c_j + Z \JJ /2 \sum_{ j  } |j|  c_{j}^\dagger c_j $;  the number of flipped spins is simply given by the displacement $|j|$ from the origin, and the second term in the Hamiltonian accounts for the corresponding exchange energy cost, i.e., the particle is bound to the origin by a linear potential due to the ``string'' of flipped spins left behind \cite{RevModPhys.66.763}. The dotted line in Fig.~\ref{fig5}c shows the mean displacement $R(t)$ of the particle in this model, which indeed coincides with the mean number of flipped spins per particle in the numerical DMFT results. As evident from a comparison of the two curves for $\JJ=0$ and  $\JJ=0.5$ (the perturbative value for the Hubbard model at $U=8$, see also Fig.~\ref{fig4}e), the effect of $\JJ$ becomes important only at longer times (when numerical data already depend on $n_\delta$), because initially the kinetic energy of the carrier is much larger than $\JJ$. In order to measure the effect of $\JJ$ on the charge-carrier interaction, one would have to reduce the number of excitations (e.g., by switching on the hopping slowly), which however makes an accurate determination of $n_\delta$ increasingly difficult.

\section{Conclusion}  

In conclusion, we have studied the short-time relaxation dynamics of the N\'eel state in the single-band Hubbard model by means of nonequilibrium DMFT, using DMRG to solve the quantum impurity model. We find qualitatively different relaxation behaviors for weak and strong interactions, separated by a crossover around $U$ $\approx 0.6\times$bandwidth: For strong interaction, local magnetic moments persist while their order is destroyed by spin-flips due to the hopping of mobile charges. The latter resembles the femtosecond carrier spin interaction which is relevant for the dynamics of photo-induced states in high-T$_c$  cuprates \cite{DalConte2015}. To demonstrate the persistence of local moments we proposed a spin precession experiment, which could be implemented similar to the proposed measurement of dynamic spin-spin correlation functions in equilibrium \cite{Knap2013}. At weak interaction, the dynamics of the N\'eel state is governed by almost conserved quasiparticles, which are also the origin for  prethermalization in nearly integrable systems \cite{Berges2004,Kollar2011,Gring2012}. In the symmetry-broken state, the breakdown of these quasiparticles away from integrability leads to a crossover from oscillatory to non-oscillatory relaxation behavior, which can provide a clear experimental signature that does now rely on a quantitative comparison to the thermal equilibrium state.

Our simulations within  DMFT are exact in the infinite dimensional limit, and it is thus 
interesting 
to compare to recent results for one dimension \cite{A.BauerF.Dorfner2015}. Similar to our results, in $d=1$ one finds a rapid saturation of the double occupancy and a slower dynamics of the order parameter at large $U$, but the decay of antiferromagnetic order is of different origin: In large dimensions, the fastest melting processes after the quench take place on the timescale of the hopping due to the strong charge-spin interaction, while the latter is absent in $d=1$ so that the dynamics happens on the timescale of the exchange interaction \cite{A.BauerF.Dorfner2015}. The quasiparticle physics at weak coupling and in the crossover regime has not been addressed in Ref.~\cite{A.BauerF.Dorfner2015}, but based on the perturbative argument given above the signatures in the off-diagonal components of the momentum distribution should persist also in lower dimensions. (Also in the paramagnetic case, a long-lived jump in the momentum distribution function is found in $d=1$ \cite{Uhrig2009,Hamerla2013, Tsuji2014} and $d=2$ \cite{Hamerla2014,Tsuji2014}.) 

Quench experiments starting from the N\'eel state have recently been performed with noninteracting fermions in one dimension \cite{PhysRevLett.113.170403}, and bosons in two-dimensions \cite{Brown2014}. Hence this setup should be a feasible approach to study fundamental aspects of the decay of antiferromagnetic long-range order in the paradigmatic Hubbard model. Moreover, on the numerical side our work emphasizes the high potential of DMRG as an impurity solver for future applications of nonequilibrium DMFT, to explore the intermediate coupling regime which is inaccessible by weak or strong coupling perturbation theory.

\section*{Acknowledgments} 
We thank E. Canovi, C. Gramsch, M. Kollar, F. Heidrich-Meisner, and U. Schollw\"{o}ck for constructive discussions. FAW acknowledges support by the research unit FOR 1807 of the DFG
and PW support from ERC FP7 starting grant No. 278023.

\appendix

\section{DMRG+DMFT setup}
\label{appDMFT}

\subsection*{General setup}

To simulate the dynamics of a lattice model which is initially in equilibrium at temperature $T=1/\beta$, we adopt the formulation of dynamical mean-field theory within the Keldysh framework (nonequilibrium DMFT), for an $L$-shaped time contour $\mathcal{C}$ which extends from initial time $t=0$ to a maximal time $t_\text{max}$ along the real time axis, back to time $0$, and along the imaginary time axis to $-i\beta$. For a general description of the formalism, as well as the notation 
and definition 
of contour-ordered functions, we refer to  Ref. \cite{Aoki2014}. In this appendix we summarize the specific setup for the quench from the N\'eel state, and the solution of the DMFT equations using DMRG. 

In DMFT, the lattice model is mapped to a set of impurity problems, one for each inequivalent lattice site $j$, with time-dependent hybridization functions $\Delta_{j\sigma}(t,t')$. The action of the impurity model is given by 
\begin{multline}
\mathcal{S}_j = -i \int_\mathcal{C} dt\, U n_{\uparrow}(t)n_{\downarrow}(t) 
\\
-i \sum_\sigma \int_\mathcal{C} dt_1dt_2\, c_\sigma^\dagger(t_1) \Delta_{j\sigma}(t_1,t_2)c_\sigma(t_2)
\label{action}
\end{multline} 
on the Keldysh contour $\mathcal{C}$, 
which yields the local contour ordered Green's function
$G_{j\sigma}(t,t')  = -i \text{Tr}[\text{T}_\mathcal{C} e^{\mathcal{S}_j} c_{\sigma}(t)c_{\sigma}^\dagger(t') ]/\mathcal{Z}$.
The hybridization function $\Delta_{j\sigma}(t,t')$,  must be defined self-consistently. For the Bethe lattice, one has \cite{RevModPhys.68.13}
\begin{equation}
\label{bethe}
\Delta_{j\sigma}(t,t') = \sum_l J(t) G_{l\sigma}(t,t') J(t')^*,
\end{equation}
where the sum runs over nearest neighbors of $j$. In the antiferromagnetic state, all sites on the $A$ and $B$ sub-lattices are equivalent, respectively. With the additional symmetry $G_{A,\sigma} = G_{B,-\sigma}$ only one impurity model must be solved with $\Delta_{\sigma}(t,t')=J_*(t)G_{-\sigma}(t,t')J_*(t')$, where we used the scaling $J(t)=J_*(t)/\sqrt{Z}$ with the coordination number $Z$. For the initial product state with $J_*(t)=0$ for $t<0$, $\Delta(t,t')=0$ if one time argument is on the imaginary branch of $\mathcal{C}$. Furthermore equivalence under a simultaneous spin and particle-hole transformation implies the symmetry
\begin{align}
\label{phs}
\Delta^>_{\sigma}(t,t') = \Delta^<_{-\sigma}(t,t')^*.
\end{align}

To compute the Green's function we follow Ref.~\cite{Gramsch2013} and map the impurity model to a time-dependent Anderson Hamiltonian 
\begin{align}
H_\text{imp} = U n_{\uparrow}n_{\downarrow} + \sum_{p\sigma} \epsilon_{p\sigma} a_{p\sigma}^\dagger a_{p\sigma}
+ \sum_{p\sigma} \big[ V_{p\sigma}(t) c_{\sigma}^\dagger a_{p\sigma} \!+\! H.c. \big]
\label{hsiam}
\end{align}
in which the impurity is coupled to $L$ bath orbitals ($p=1,...,L$).  The parameters $V_{p\sigma}(t)$ and $\epsilon_p$ are determined such that the local Green's functions obtained from \eqref{action} and \eqref{hsiam} are identical. As derived in Ref.~\cite{Gramsch2013}, for $J_*(t<0)=0$, one can choose $V_{p\sigma}(t)=0$ for $t<0$, and the mapping condition is satisfied by (assuming $L$ even)
\begin{align}
\label{lesfit}
\Delta^<_{\sigma}(t,t') &= i \sum_{p=1}^{L/2} V_{p\sigma}(t)V_{p\sigma}^*(t')
\\
\label{gtrfit}
\Delta^>_{\sigma}(t,t') &= -i \sum_{p=L/2+1}^L V_{p\sigma}(t)V_{p\sigma}^*(t'),
\end{align}
where $\epsilon_{p\sigma}=0$, and the bath orbitals $p=1...L/2$ and $p=L/2+1...L$ are initially doubly occupied and empty, respectively. Equations \eqref{lesfit} and \eqref{gtrfit} are solved by a Cholesky fit of the real-time matrix $\Delta(t,t')$, which quickly converges for small times with the number of bath orbitals required \cite{Balzer2014}. Due to the symmetry \eqref{phs} we use  
\begin{align}
V_{p,-\sigma}(t) = V_{L/2+p,\sigma}(t)^* \,\,\,\text{~for~} p\le L/2.
\end{align}
The impurity site is initially occupied with a spin $\sigma=\uparrow$ (for a site on the $A$ sublattice), i.e., the initial state for the impurity model is a product state
$ |\Psi_{\text{imp},A}\rangle =  c_{\uparrow}^\dagger \prod_{i=1}^{L/2} a_{p\uparrow}^\dagger a_{p\downarrow}^\dagger |0\rangle$, and the Green's function is obtained by solving 
\begin{align}
G_{A,\sigma}^<(t,t') 
&= i \langle\Psi_{\text{imp},A}|  c_{\sigma}^\dagger(t') c_{\sigma}(t) |\Psi_{\text{imp},A}\rangle,
\\
G_{A,\sigma}^>(t,t') 
&= -i \langle\Psi_{\text{imp},A}|  c_{\sigma}(t) c_{\sigma}^\dagger(t') |\Psi_{\text{imp},A}\rangle,
\end{align}
where time evolution is determined by \eqref{hsiam}. We use a Krylov time-propagation for matrix product states \cite{Wolf2014} with up to $L=24$ bath orbitals. 

\subsection*{Inhomogeneous setup} 

For the inhomogeneous setup we assume that in the initial state on the lattice the spin at one site $o$ of the lattice is flipped in the $x$ direction. Without loss of generality we assume that $o$ is on the $A$ sub-lattice. From the self-consistency equation \eqref{bethe} one can see that the hybridization on all other sites differs from the homogeneous case only in order $1/Z$, i.e. for $Z\to\infty$ the back-action of the probe site on the rest of the lattice can be neglected. On the probe site we solve an impurity problem with the same (nonequilibrium) hybridization function $\Delta_{A}$ as on all remaining $A$-sites, i.e., an impurity problem \eqref{hsiam} wit the same parameters $V_{p\sigma}$, but with a different initial state,
\begin{align}
|\Psi_{\text{imp},o}\rangle =  
\Big(
c_{\uparrow}^\dagger \prod_{i=1}^{L/2} a_{p\uparrow}^\dagger a_{p\downarrow}^\dagger |0\rangle
+
c_{\downarrow}^\dagger \prod_{i=1}^{L/2} a_{p\uparrow}^\dagger a_{p\downarrow}^\dagger |0\rangle
\Big)/\sqrt{2}.
\end{align}

\subsection*{Observables} 

Local observables 
$
\langle
\mathcal{O}_j(t)
\rangle
\equiv
\langle \Psi_{\text{imp},j} 
|
\mathcal{O}(t)
|
\Psi_{\text{imp},j}\rangle$
are directly measured in the impurity model ($j=o,A$),
in particular 
the density
$\mathcal{O} 
\equiv
n_{\sigma}
$,
the double occupancy 
$\mathcal{O} 
\equiv
n_{\uparrow}
n_{\downarrow}
$,
and the spin 
$\mathcal{O} 
\equiv 
S_{\alpha=x,y,z}  = \tfrac12 \sum_{\sigma\sigma'} c_{\sigma}^\dagger \bm \tau_\alpha c_{\sigma'} $
($\bm \tau_\alpha$ are the Pauli matrices).

In the translationally invariant case (no probe site), we also determine diagonal and off-diagonal components of the momentum occupations $n(\epsilon,t)$ and $\pi(\epsilon,t)$, which are obtained from the momentum resolved Green's function (for the definition of $k$ and $\bar k$, see the main text)
\begin{align}
\bm G_{\epsilon_k}(t,t') = 
\begin{pmatrix}
-i \langle T_\mathcal{C} c_{k} (t) c_{k}^\dagger(t')  \rangle
&
-i \langle T_\mathcal{C} c_{k} (t) c_{\bar k}^\dagger(t')  \rangle
\\
-i \langle T_\mathcal{C} c_{\bar k} (t) c_{k}^\dagger(t')  \rangle
&
-i \langle T_\mathcal{C} c_{\bar k} (t) c_{\bar k}^\dagger(t')  \rangle
\end{pmatrix}.
\end{align}
(Here and in the following, bold-face quantities denote $2\times2$ matrices and we omit spin indices for simplicity). The self-energy is local in space but depends on the sub-lattice and spin; in the $k$,$\bar k$ representation it thus assumes the ($2\times2$) form
\begin{multline}
\bm \Sigma (t,t') = 
\tfrac12
[\Sigma_A(t,t')+\Sigma_B(t,t')] \bm 1
\\+
\tfrac12
[\Sigma_A(t,t')-\Sigma_B(t,t')] \bm\tau_x,
\end{multline}
so that $\bm G_\epsilon$ is obtained from the lattice Dyson equation 
$\bm G_{\epsilon} = (i\partial_t + \mu - \bm \epsilon - \bm \Sigma)^{-1} $,
where the dispersion in the $k$,$\bar k$ representation reads $ \bm \epsilon =  \epsilon \bm \tau_z $ because 
$\epsilon_k=-\epsilon_{\bar k}$. The components $\Sigma_{j}$ of the self-energy  ($j=A,B$) are obtained from the  impurity Dyson equation $(i\partial_t + \mu - \Delta_{j} - \Sigma_{j})^{-1} = G_{j}$. In praxis, we solve an integral equation $G_{j} = Z_{j} + Z_{j} \ast \Delta_{j} \ast G_{j}$ for $Z_{j} = (i\partial_t + \mu  - \Sigma_{j})^{-1}$. We then have  $\bm Z=(i\partial_t + \mu  - \bm \Sigma)^{-1} = \tfrac12(Z_A+Z_B)\bm 1 + \tfrac12(Z_A-Z_B) \bm \tau_x$, and $\bm G_\epsilon$ is obtained from the integral equation  $\bm G_{\epsilon} = \bm Z + \bm Z \ast \bm \epsilon \ast \bm G_{\epsilon}$.

\section{Mobile carrier density in the excited state.}
\label{appndoublon}

\begin{figure}[tbp]
\begin{center}
\includegraphics[angle=0, width=\columnwidth]{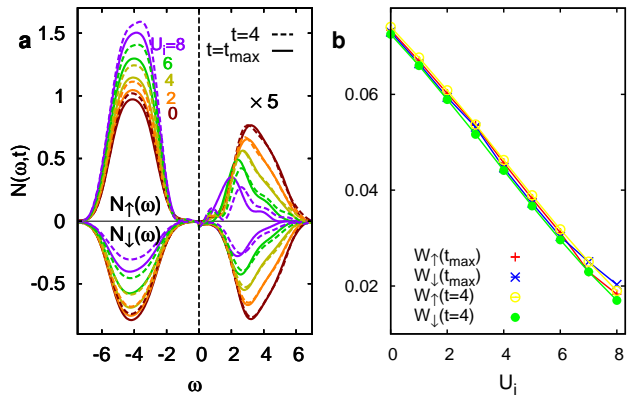}%
\caption{
{\bf a)} Occupied density of states $N_{\sigma}(\omega,t)$ after the quench 
$t<0:$ $v=0$ (N\'eel state);
$0\le t <0.5$: $v=1$, $U=U_i$;
$t\ge 0.5$: $v=1$, $U=8$; for  various values of $U_i$; $N_{\uparrow}(\omega,t)$ and  $N_{\downarrow}(\omega,t)$ refer to majority and minority spin, respectively, and the upper Hubbard band is scaled  by a factor $5$. Dashed and solid lines denote time $t=4$ and the largest simulation time $t=t_\text{max}$, respectively. {\bf b)} Integrated weight in the upper Hubbard band (red crosses and blue stars).}
\label{fig6}
\end{center}
\end{figure} 

In the Mott insulating phase of the Hubbard model, a well-defined measure for the number of doublon or hole carriers is given by the total occupied spectral weight in the upper Hubbard band and the total unoccupied weight in the lower Hubbard band, respectively. The double occupancy, in contrast, depends on virtual charge fluctuations which are nonzero also in the insulating ground state. Specifically, we define the occupied density of states as the partial Fourier transform $N_\sigma(t,\omega) = \text{Im} \int_0^t ds \exp(-s^2/2\delta^2) \exp[-i s\omega] G_\sigma^<(t-s,t)$ where $G_\sigma^<(s,s')=i\langle c^\dagger_\sigma(s') c_\sigma(s) \rangle $ is the local Green's function, and $\delta=1.5$ ensures a smooth cutoff (which does not influence the results unless its inverse width is longer than the inverse of the gap). The spectrum  $N_\sigma(\omega,t)$ is plotted in Fig.~\ref{fig6}a for two different times, for the same quench parameters as in Fig.~\ref{fig5} of the main text. The right panel shows the integrated density $W_\sigma(t) = \int_0^\infty d\omega N_\sigma(\omega,t)$. While the weight in the upper and lower band differs considerably between majority and minority spin, the integrated  weight $W_\sigma(t)$ reflects the doublon density and is thus independent of $\sigma$. It is interesting to point out that as a function of time spectral weight is both redistributed between the lower Hubbard bands of the two spin components (which reflects the decay of the N\'eel order), and within the upper Hubbard band (which reflects the change of the kinetic energy of the doublons), while the total weight in the upper band is roughly constant (see, e.g., Ref.~\cite{Werner2012}). For the analysis in the main text, we take $n_\delta = (W_{\uparrow}(t_\text{max})
+W_{\downarrow}(t_\text{max}))/2$.

\section{Solution for $U=0$}
\label{appU0}

For $U=0$ the time evolution of the N\'eel state on the Bethe lattice can be obtained analytically by solving the Heisenberg equations of motion for the $c$-operators, which provides a good check for the numerical implementation.  For completeness, we provide this solution in the following. We choose site $0$ to be the origin of the Bethe lattice, which is on the $A$ sub-lattice without loss of generality. One can map the solution of equations of motion on the Bethe lattice to a one-dimensional semi-infinite chain by introducing operators which are invariant under all permutations of the branches of the  Bethe lattice \cite{Mahan2001},
\begin{align}
\label{Cn}
C_{n} = \frac{1}{\sqrt{Z_n}}\sum_{i:|i-0|=n} c_i,
\end{align}
where $Z_n = \sum_{i:|i-0|=n}$ is the number of sites on the $n$-th nearest neighbor shell.
Then the action of the Hamiltonian is determined by 
$[H,C_j]=-\sum_{i=0}^\infty h_{ji}C_i$, with 
\begin{align}
h =
\begin{pmatrix}
0 & 1 & 0 & 0 & \cdots 
\\
1 & 0 & 1 & 0 & 
\\
0 & 1 & 0 & 1 &  
\\
\vdots & & & & \ddots
\end{pmatrix}.
\end{align}
Hence, eigenvectors for the eigenvalue $\epsilon$ satisfy the equation
\begin{align}
&\phi(\epsilon)_0 = 1 \\
&\phi(\epsilon)_1 = \epsilon \\
&\epsilon \phi(\epsilon)_{n} = \phi(\epsilon)_{n+1}+ \phi(\epsilon)_{n-1},
\end{align}
and are thus given by the Chebychev polynomials of second kind \cite{Abramowitz1965},
$\phi(\epsilon)_{n} = U_n(\epsilon/2)$ for $-1\le \epsilon/2 \le 1$. The $U_n$ can be conveniently written as 
\begin{align}
U_n(\cos(\theta)) = \frac{\sin[(n+1)\theta]}{\sin(\theta)}, 
\end{align}
from which one  can also see the orthogonality
\begin{align}
\int_{-1}^1 dx \,w(x)\, U_n(x) U_m(x) =\delta_{mn}
\end{align}
with $w(x)=\frac{2}{\pi}\sqrt{1-x^2}$. Thus the solution of the Heisenberg equations of motion for the local $c$ operator \eqref{Cn}
\begin{align}
\frac{d}{dt} C_0(t) = i [H,C_0(t)], \,\,\, C_0(0)=C_0,
\end{align}
is given by
\begin{align}
&
C_0(t) = \sum_{n=0}^\infty \psi_n(t) C_n,
\\
&
\psi_n 
=
\int_{-1}^1 dx \,w(x)\,  e^{-i2xt} U_n(x).
\end{align}
This can be transformed to
\begin{align}
\psi_n 
&=
\frac{2}{\pi} \int_{-1}^1 d(\cos(\theta)) \,  \sin(\theta)  e^{-i2\cos(\theta)t} \frac{\sin[(n+1)\theta]}{\sin(\theta)}
\nonumber\\
&=
\frac{2}{\pi} \int_{0}^\pi d\theta  \sin[(n+1)\theta] \frac{1}{2it} \partial_\theta e^{-i2\cos(\theta)t} 
\nonumber\\
&=
\frac{i}{\pi t} \int_{0}^\pi d\theta   e^{-i2\cos(\theta)t} \partial_\theta \sin[(n+1)\theta] 
\nonumber\\
&=
\frac{i(n+1)}{\pi t} \int_{0}^\pi d\theta   e^{-i2\cos(\theta)t} \cos[(n+1)\theta] 
\nonumber\\
&=
\frac{i(n+1)}{\pi t} \int_{0}^\pi d\theta   e^{i2\cos(\theta)t} \cos[(n+1)\theta] (-1)^{n+1}
\nonumber\\
&=
(- i)^n (n+1) \frac{J_{n+1}(2t)}{t}.
\end{align}
The second to last line is a variable transformation $\theta\to\pi-\theta$, and in the last line we have used the integral representation of the Bessel function \cite{Abramowitz1965},
\begin{align}
J_n(z) = \frac{(-i)^n}{\pi} \int_{0}^\pi d\theta \cos(n\theta) e^{iz\cos(\theta)}.
\end{align}

The explicit form of the $C$-operators can be used to obtain local observables
\begin{align}
\langle 
\Psi
|
c_{\sigma}^\dagger(t)
c_{\sigma'}(t)
|
\Psi
\rangle
&=
\sum_{n,m}
\psi_n^*(t)
\psi_m(t)
\langle \Psi |
C_{n\sigma}^\dagger
C_{m\sigma'}
|
\Psi
\rangle
\\
&=
\sum_{n}
|\psi_n(t)|^2
\langle \Psi |
C_{n\sigma}^\dagger
C_{n\sigma'}
|
\Psi
\rangle,
\end{align}
where the expectation values are simple initial state values. We start by evaluating the time evolution of the magnetic order,
$n_{\uparrow}-n_{\downarrow}$, at site $0$, in the classical Neel state. For the latter we have
\begin{align}
\langle \Psi_\text{N\'eel} |
C_{n\uparrow}^\dagger
C_{n\uparrow}
-
C_{n\downarrow}^\dagger
C_{n\downarrow}
|
\Psi_\text{N\'eel}
\rangle
=
(-1)^n,
\end{align}
and hence 
\begin{align}
\langle \Psi_\text{N\'eel} |
n_{\uparrow}
(t)
-
n_{\downarrow}(t)
|
\Psi_\text{N\'eel}
\rangle
=
\sum_{n=0}^\infty
(-1)^{n}
\frac{(n+1)^2 J_{n+1}(2t)^2}{t^2},
\label{afm1}
\end{align}
(where the summation index has been shifted by one).
We can now use Gegenbauers addition theorem for Bessel functions \cite{Abramowitz1965}
to obtain the final result
\begin{align}
\langle \Psi_\text{N\'eel} |
n_{\uparrow}
(t)
-
n_{\downarrow}(t)
|
\Psi_\text{N\'eel}
\rangle
=\frac{J_1(4t)}{2t},
\end{align}
which fits the numerics.

Next we compute site $0$ expectation values on the probe site. Now the initial state is a superposition
\begin{align}
&|\Psi\rangle  
=  (|\Psi_\text{N\'eel},\uparrow\rangle  + |\Psi_\text{N\'eel},\downarrow\rangle)/\sqrt{2}
\\
&
|\Psi_\text{N\'eel},\sigma\rangle 
=
c_{0,\sigma}^\dagger c_{0,\uparrow}|\Psi_\text{N\'eel}\rangle.
\end{align}
We evaluate the cross-spin expectation values
\begin{align}
\langle 
S^{+}_0(t)
\rangle 
&=
\langle 
\Psi
|
c_{0\uparrow}^\dagger(t) c_{0\downarrow}(t)
\Psi
\rangle
\nonumber\\
&=
\sum_{n}
|\psi_n(t)|^2
\langle \Psi |
C_{n\uparrow}^\dagger
C_{n\downarrow}
|
\Psi
\rangle.
\end{align}
Spin-flip expectation values are only non-zero in the initial state at site $0$, where we have 
\begin{align}
\langle 
S^{+}_0(t)
\rangle 
&=
|\psi_0(t)|^2
\langle \Psi |
c_{0\uparrow}^\dagger
c_{0\downarrow}
|
\Psi
\rangle
\\
&
=
\frac{J_1(2t)^2}{2t^2}.
\end{align}
Hence $S^{+}_0(t)$ is purely real, so that the dynamics is entirely longitudinal in the $S_x$-$S_y$-plane,
\begin{align}
\langle 
S^{x}_0(t) 
\rangle
&= \frac{J_1(2t)^2}{2t^2}. 
\\
\langle 
S^{y}_0(t) 
\rangle
&= 0.
\end{align}


\begin{thebibliography}{59}
\bibitem{footnote1}
In a cold-atom experiment, one could switch off the hopping slowly compared to the Mott gap, in order to adiabatically reduce the quantum fluctuations.
\end{thebibliography}

\end{document}